\documentclass[prl,superscriptaddress,twocolumn]{revtex4-2}
\usepackage[utf8]{inputenc}
\newcommand{\ud}{\mathrm{d}}
\usepackage{amsmath}
\usepackage{amssymb}
\let\lambda\lambdaup
\allowdisplaybreaks
\usepackage{graphicx}
\graphicspath{{fig/}}
\usepackage{xcolor}
\usepackage[colorlinks=true,linkcolor=blue,citecolor=blue,urlcolor=blue]{hyperref}

\begin{document}

\title{High-temperature transport and polaron speciation in the anharmonic Holstein model}

\author{Attila Szab\'{o}}
\address{Rudolf Peierls Centre for Theoretical Physics, University of Oxford, Oxford OX1 3PU, UK}
\address{ISIS Facility, Rutherford Appleton Laboratory, Harwell Campus, Didcot OX11 0QX, UK}
\author{S. A. Parameswaran}
\address{Rudolf Peierls Centre for Theoretical Physics, University of Oxford, Oxford OX1 3PU, UK}
\author{Sarang Gopalakrishnan}
\affiliation{Department of Physics, The Pennsylvania State University, University Park, PA 16802, USA}

\date{\today}

\begin{abstract}

We study the finite-temperature transport of electrons coupled to anharmonic local phonons. Our focus is on the high-temperature incoherent regime, where controlled calculations are possible both for weak and strong electron--phonon coupling. At strong coupling, the dynamics is described in terms of a multiple-species gas of small polarons: this emergent `speciation' is driven by energy conservation. We explicitly compute the dc and ac response in this regime. We discuss the breakdown of the polaron picture, and the onset of localization, in the limit where  phonons become quasistatic.

\end{abstract}

\maketitle

Phonons impede transport in metals and facilitate it in localized insulators, in both cases by decohering electronic degrees of freedom. Acoustic phonons, in particular, can act as an effective bath for electrons, allowing incoherent hopping transport in otherwise localized systems~\cite{Miller1960,Shklovskii1984,Pollak2012,Fleishman1980}. More generally, however, phonons are quantum degrees of freedom that can themselves localize~\cite{PhysRevX.3.021017, DeRoeck2014, Banerjee2016}, or otherwise fail to act as an effective bath. It is well-established, for example, that optical phonons suppress transport by binding to electrons and forming heavy composite excitations, or \emph{polarons}~\cite{alexandrov1996polarons,Franchini2021}.

\begin{figure}
    \centering
    \includegraphics{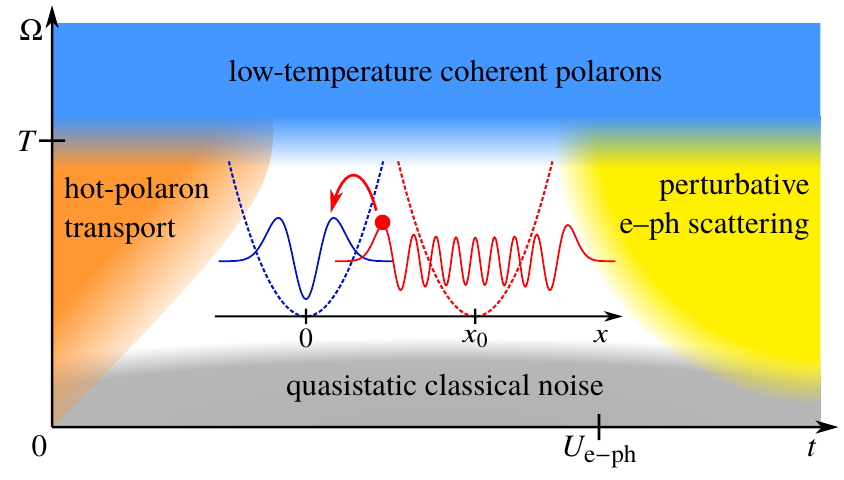}
    \caption{Schematic transport regimes vs. hopping $t$ and phonon frequency $\Omega$ for fixed temperature $T$ and electron-phonon coupling $U_\text{e--ph}$. The four coloured regions  admit controlled calculations; the white area in the centre does not.
    \textit{Inset:} Facilitated dynamics emerge because rare, highly excited phonon states on occupied sites (red) have exponentially larger overlap with thermal states on empty sites (blue), making the associated polarons much more mobile.}
    \label{fig1}
\end{figure}

Although the transport properties of polarons at low temperatures have been extensively studied~\cite{alexandrov1996polarons,Franchini2021,Han2020,Huang2021,Costa2020}, the \emph{high-temperature} limit (relevant, e.g., to disorder-free MBL~\cite{DeRoeck2014, Gopalakrishnan2020}) has received less attention. Recently, a number of works have proposed that such hot, slowly fluctuating modes are relevant for transport in a number of materials such as organic semiconductors and halide perovskites~\cite{fratini2016transient, delacretaz2017bad, fratini2020charge, lacroix2020modeling, fratini2021displaced}.
However, these works do not account for the quantum dynamics of the phonons (but see Ref.~\cite{lai2020absence}). The present work addresses this gap. We characterize transport in the canonical Holstein model~\cite{Holstein1959}, through controlled estimates in three overlapping limits: weak or strong electron--phonon coupling at any temperature, and slow ``quasiclassical'' phonons for any electron-phonon coupling. 
We model the phonons as anharmonic; we argue that this anharmonicity drastically affects high-temperature response.
In the strong-coupling regime, which we address in most detail, anharmonicity
gives rise to an emergent conservation law by reducing the number of available resonances~\cite{DeRoeck2014}:
electrons can only hop to empty sites while preserving the phonon number associated with themselves and the empty site, dividing electrons and holes into many subspecies labelled by the associated phonon number.
While these  `speciated' polarons all remain mobile, we find that low-frequency transport is dominated by facilitated processes driven by a small fraction of inherently fast high-phonon-number electrons and holes, which locally make their slow counterparts mobile.
Similar facilitated processes drive dynamical heterogeneity in glasses~\cite{Biroli2013,Berthier2011Book,Berthier2011RMP,Chandler2010} and disorder-free MBL systems~\cite{Lan2018,Gopalakrishnan2020}.
At high frequencies, this facilitated dynamics gives way to resonant processes where an electron ``rattles'' between two sites, yielding an ac conductivity with anomalous frequency dependence. Our work  illustrates that unusual transport phenomena akin to those seen in the MBL phase can emerge in extreme regimes of even well-studied models.

\textit{Model.---}%
Our focus is finite-temperature transport in the anharmonic Holstein model, with Hamiltonian 
\begin{equation}
    H = \sum_{ij} t_{ij} c^\dagger_i c_j + \text{h.c.} + \sum_i \left(\frac{p_i^2}{2m} + V(x_i) -  \alpha n_i x_i\right).
    \label{eq: Hamiltonian before polaron}
\end{equation}
$c^\dagger_i$ creates a spinless fermion~%
\footnote{We refer to these fermions as `electrons;' treating them as spinless is physical, e.g., if strong repulsive interactions prevent doubly occupied sites~\cite{Han2020}.}
on site $i$; $n_i = c^\dagger_i c_i$; and $p_i, x_i$ are the momentum and position of the anharmonic oscillator at site $i$, with potential  $V(x_i) = \frac{1}{2} k x_i^2 + c x_i^4 + \ldots$ (we require $c \neq 0$) and frequency  $\Omega \equiv \sqrt{k/m}$. 
Most of our results are geometry-independent, but for concreteness we consider a one-dimensional (1D) chain with beyond-nearest-neighbor hopping, which  has the \emph{generic} feature that electrons can traverse nontrivial loops. 
The strength of electron--phonon coupling is captured by the shift $U_\mathrm{e-ph}$ in the phonon ground-state energy due to an onsite electron; for weak anharmonicity ($c \langle x^2 \rangle \ll k$ for states of interest), $U_{\mathrm{e-ph}} \approx \alpha^2/(2k)$.

The four characteristic energy scales---$t, \Omega, U_{\mathrm{e-ph}}$ and the temperature $T$---define a three-dimensional parameter space. Fig.~\ref{fig1} shows a cut through this phase diagram at a fixed $T \ll U_{\mathrm{e-ph}}$; the motivation for this choice will be explained below. We focus on the behavior at temperatures where the phonons are thermally occupied, i.e., $\Omega \ll T$, as the opposite case has been extensively discussed~\cite{alexandrov1996polarons,Franchini2021,Han2020,Huang2021,Costa2020}. Transport can be discussed in a controlled way in three sectors of this space: (i)~the weak-coupling regime $U_{\mathrm{e-ph}} \ll t$, (ii)~the quasistatic regime $\Omega \ll U_{\mathrm{e-ph}}, t$, and (iii)~the weak-hopping regime $t \ll \Omega, U_{\mathrm{e-ph}}$. Some of these regimes are amenable to standard approaches and we will simply sketch the main ideas; we will dwell on the case $t \ll \Omega \ll T \ll U_{\mathrm{e-ph}}$, as it exhibits some qualitatively new features.

\emph{Weak-coupling regime}.---%
In the weak-coupling regime, we neglect the phonons to leading order and compute the conductivity by diagrammatic perturbation theory in the electron--phonon coupling around the free-electron Bloch eigenstates. 
The main difference from the diagrammatic perturbation theory for disordered systems is that scattering off phonons can lead to energy relaxation.
\emph{Anharmonicity} is crucial for this broadening: if the phonons were perfectly harmonic, they would only mix states separated by integer multiples of $\Omega$, and full energy relaxation would not occur. The broadening due to anharmonicity scales as $w \equiv cT/\sqrt{k^3m}$.
The existence of this relaxation channel is one crucial distinction between our treatment and that of Ref.~\cite{lai2020absence}.

At the diffusive level, this approach would give a momentum relaxation time $\tau \sim t/(U_{\mathrm{e-ph}} T)$, and thus a diffusion constant $D \sim t^3/(U_{\mathrm{e-ph}} T)$. This perturbation theory breaks down, however, when $\Omega \to 0$ in dimensions $d < 3$. Since the phonons are effectively static on timescales $\alt 1/w$, the diffusion constant has a singular weak localization (WL) correction~\cite{RevModPhys.57.287}. In 2D, this only matters for exponentially small $w$; in 1D, however, the WL correction becomes important on the scale of the mean free time, i.e., when $w t \approx U_{\mathrm{e-ph}} T$. To address physics  below this scale, we turn to the quasistatic limit.

\emph{Quasistatic limit}.---%
We now consider the limit $\Omega\to0$. Specifically, we take $m\to\infty$, i.e., make $x$ a static distortion, reducing $H$ to an Anderson localization problem of electrons in a random potential drawn from the Gaussian distribution $P(x_i) \propto \exp(-k x_i^2/(2T))$. 
In $d=1,2$, the electrons are always localized; in 3D, there is a localization transition tuned by the  `disorder strength' $T/k$. 

We now consider small nonzero $\Omega$. A natural semiclassical approach is to initialize the oscillators in a typical thermal configuration and let them evolve classically. Consider weakly localized orbitals, $\xi \gg 1$, so that each electron is spread out over $\xi^d$ sites. 
The instantaneous phonon interaction energy of each electron is thus the average of $\sim\xi^d$ terms that all precess at different on-site frequencies $\omega_i \in (\Omega, \Omega + w)$.
One can regard this as noise with bandwidth $w$. Two types of behavior are possible, depending on how $w$ compares with the level spacing within a localization volume, $\delta_\xi \sim t/\xi^d$. 
If $w > \delta_\xi$, the noise is able to drive transitions within a localization volume, and localization is irrelevant. (In 1D, $w \sim \delta_\xi$ reproduces our previous estimate from the weak-coupling side.) 
On the other hand, if $w < \delta_\xi$, transport occurs via Landau--Zener transitions. On a timescale $1/w$, the local potential landscape completely rearranges,
driving resonant transitions between localized states and thus to transport. Up to possible logarithmic corrections, transport in this regime is controlled by the scale $w$, and $D \sim \xi^2 w$. 
Indeed, if the phonons are harmonic ($w=0$), diffusion may never occur~\cite{lai2020absence}.

\emph{Strong-coupling regime}.---%
We finally turn to the strong-coupling regime. Here, one can work perturbatively in $t$ while treating the other couplings exactly.
While there are several possible hierarchies of couplings in this regime, we will anticipate that interesting physics occurs for the hierarchy of couplings $t \ll \Omega \ll T \ll U_\mathrm{e-ph}$, and comment briefly on the other cases later.

At zeroth order in $t$, each site has some sharp phonon occupation number drawn from the Gibbs distribution. 
At electron-occupied sites, the effective oscillator potential $kx^2 + c x^4 + \ldots - \alpha x$ is minimized at $x_0 \approx \alpha/k$. 
(To retain simple closed-form expressions, we work in the limit of relatively weak anharmonicity, i.e., the frequencies of shifted and unshifted oscillators differ by much more than $t$, but $k \gg cx_0^2$, so the two  are of the same order of magnitude. The qualitative picture remains the same in the strongly anharmonic limit.) 
Strong electron--phonon coupling is defined by the condition $x_0 \gg l_\Omega$, where $l_\Omega$ is the size of the oscillator ground state: we define $A\equiv x_0/l_\Omega\approx \sqrt{U_{\mathrm{e-ph}}/\Omega}$. 
We will take $A \gg \sqrt{T/\Omega}$, so the shifted and unshifted oscillator ground states (corresponding respectively to occupied and empty sites) overlap only in their Gaussian tails (Fig.~\ref{fig1}, inset). 

We now introduce hopping at lowest order in perturbation theory. Suppose there is an electron on a site with $n$ phonons, and its empty neighbor has $m$ phonons. Generically, the only resonant hopping process is one in which the electron carries along its phonon cloud, i.e., the empty $m$-photon site and the occupied $n$-photon site swap places.  
Thus, each electron (and each empty site) is ``bound'' to its $n$-phonon cloud; moreover, when it hops, it exchanges occupation numbers with the destination site. Thus the electron leaves a trail of altered occupation numbers as it moves. This `polaron speciation', a consequence of energy conservation in the strong-coupling limit, has particularly important effects when electrons traverse closed loops in configuration space. 
For instance, if an electron hops along a loop $a \to b \to c \to d \to a$, it permutes the occupation numbers of sites $b, c, d$ along the cycle $(bcd)$, whereas the reverse hop permutes the occupation numbers as $(bdc)$. The persistence of this which-way information suppresses interference effects, making  transport incoherent even in the absence of an external bath.   

The transition matrix element between an $n$-phonon electron and an $m$-phonon empty site is $t |D_{mn}|^2$, where the Franck--Condon (FC)  factor $D_{mn}$ is the overlap between the shifted $n$-phonon state and the unshifted $m$-phonon state. 
If $T\ll U_\mathrm{e-ph}$, typical  phonon states (i.e., those with $n,m\sim T/\Omega\ll A^2$) on occupied and empty sites only overlap in their Gaussian tails, thus the FC factor for most hopping processes is exponentially suppressed.
A small number of electrons and empty sites (which we term ``fast''), however, has $n,m\approx A^2$: as these phonon wave functions overlap with the classically allowed regions of typical states, the associated   FC factors are $O(1)$, making them exponentially more mobile than a typical electron or empty site. 
We now explore the effect of this dynamical heterogeneity on ac transport.

\emph{Two-site transport}.---%
The first contribution is from coherent processes between two degenerate states coupled by a hopping matrix element, which can be calculated using the Kubo formula.
As noted before, we can restrict our attention to processes where a electron hops from site $i$ to site $j$ and the sites exchange their phonon content. The eigenstates of this two-site system are $(|\dot m n\rangle \pm |n\dot m\rangle)/2$ with energies $\mp t_{ij}|D_{mn}|^2$. (Dots above phonon numbers indicate electron occupation.) The one-dimensional conductivity is now given by the Kubo result
\begin{align}  
    \sigma_{ij}(\omega) &= 4\pi\beta L \frac{e^2}{\hbar} \rho(1-\rho)(1-e^{-\beta\hbar\Omega})^2 \times   \label{eq: two state Kubo formula}\\*
    & \sum_{n,m=0}^\infty e^{-\beta\hbar\Omega(n+m)}t_{ij}^2 |D_{mn}|^4  \delta(\hbar\omega - 2t_{ij} |D_{mn}|^2),\nonumber
\end{align}
where $\rho$ is the density of electrons [$2\rho(1-\rho)$ is the probability of having precisely one electron on the two sites], $(1-e^{\beta\hbar\Omega})^{-1}$ is the phonon partition function of a single site, and the length scale $L$ is on the order of the lattice spacing, since the phonon-induced disorder decouples hopping processes between different pairs of sites.

\begin{figure}
    \centering
    \includegraphics{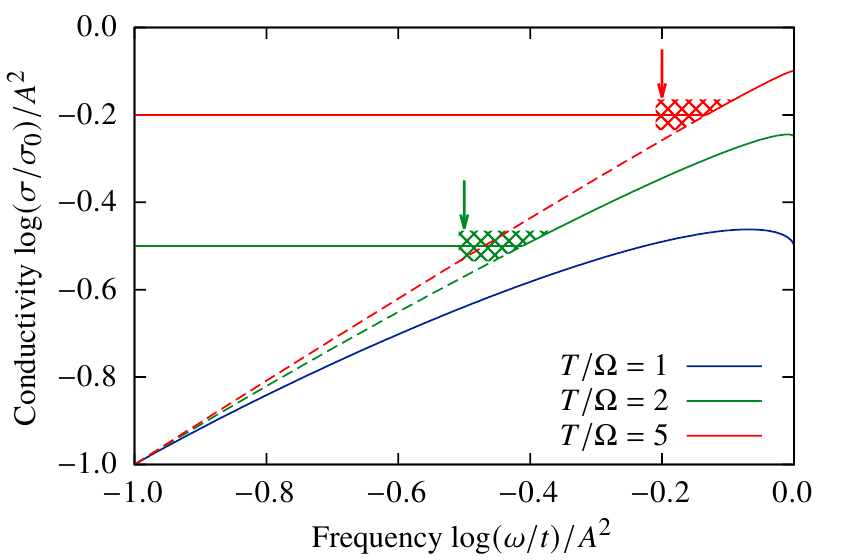}
    \caption{Log-log plot of ac conductivity in the anharmonic Holstein model.
    At frequencies below $\Gamma_0\approx te^{-U_\mathrm{eff}/T}$ (arrows), transport is dominated by facilitated processes due to incoherent hopping of ``fast'' electrons and holes, with frequency-independent conductivity $\sigma_0 e^{-U_\mathrm{eff}/T}$.
    At high frequencies, this gives way to two-site resonant processes (extrapolated to low frequencies with dashed lines), with conductivity that starts off linear in frequency, saturates, then reduces to $\sigma_0 e^{-U_\mathrm{eff}/2T}$ at the highest resonantly available frequency, $\omega=2t$. 
    At frequencies  $\omega\sim\Gamma_0$ (shaded), other processes (e.g., multisite coherent hopping) may also be relevant. }
    \label{fig: conductivity}
\end{figure}

We now evaluate~\eqref{eq: two state Kubo formula} in the limit $\hbar\Omega\ll T\ll U_\mathrm{e-ph}$.
Asymptotically in the strong-coupling limit, we have
\begin{equation}
    \log |D_{nm}| = -A^2 F\left(\frac{n}{A^2}, \frac{m}{A^2}\right),
    \label{eq: Franck-Condon asymptote}
\end{equation}
where $F(x,y)$ is a universal scaling function~\cite{supplement}. 
At high temperatures, we replace the sum over $n,m$ by an integral, which we can treat in the saddle-point approximation due to the exponential suppression of~\eqref{eq: Franck-Condon asymptote}. 
At all frequencies, $\sigma(\omega)$  is dominated by $n=m=\bar{n}$ such that $2t_{ij}|D_{\bar{n}\bar{n}}|^2= \hbar\omega$~\cite{supplement}. 
Fig.~\ref{fig: conductivity} shows the asymptote of $\sigma(\omega)$ for several temperatures.
At low frequencies, $\sigma\propto\omega$ due to the $t_{ij}^2|D_{mn}|^4\propto \omega^2$ factor in~\eqref{eq: two state Kubo formula} and a density-of-states factor scaling as $\omega^{-1}$ that arises from the on-shell constraint and the exponential scaling of~\eqref{eq: Franck-Condon asymptote}.
At higher frequencies, the Boltzmann-factor term in~\eqref{eq: two state Kubo formula} changes this scaling: 
the FC factor saturates once the classically allowed regions of initial and final phonon states overlap, causing $\sigma(\omega)$ to saturate and briefly turn decreasing near the highest possible resonance frequency, $\omega_\mathrm{max}=2t_{ij}/\hbar$.
However, this effect becomes less pronounced as temperature increases.
At the highest frequencies, $\sigma(\omega_\mathrm{max}) \approx \sigma_0 e^{-U_\mathrm{e-ph}/2T}$, where $\sigma_0 \sim \beta L e^2\rho(1-\rho) (1-e^{-\beta\hbar\Omega})^2t_{ij}/\hbar$ combines all dimensionful prefactors in~\eqref{eq: two state Kubo formula}. (We omit dimensionless prefactors from preexponential terms of saddle-point integrals.)

\textit{Incoherent multi-site processes.---}%
Such coherent processes do not survive to arbitrarily low frequencies: high-phonon-number, ``fast'' electrons or holes reach the sites at a characteristic rate and ``scramble'' their contents, disrupting coherent dynamics.
Therefore, the dynamics of the vast majority of electrons is dominated by hopping processes facilitated by rare, fast holes (and vice versa), making them incoherent and diffusive.

To estimate the rate of such processes, we first calculate the hopping rate of an electron at $i$ with phonon number $n\gg T/\Omega$. Consider a hole at $j$ with phonon number $m$: 
ignoring the rest of the system, the states $|\dot n m\rangle$ and $|m\dot n\rangle$ hybridise and split by energy $2t_{ij}|D_{nm}|^2$, which sets the frequency of Rabi oscillations between the split states. 
However, once the electron has moved to $j$, it is able to move on to other empty sites in addition to moving back to $i$, which disrupts the cycle of Rabi oscillations. Therefore, it is better to think of this process as stochastic hopping at rate $\Gamma \sim t_{ij} |D_{mn}|^2$. 
Unlike Fermi's golden rule, the rate is proportional to the matrix element rather than its square: 
intuitively, the splitting between $|\dot n m\rangle \pm |m\dot n\rangle$ defines  a ``density of states'' inversely proportional to the matrix element.

We assume that the hopping matrix element decays exponentially with distance:
$t_{ij} = t e^{-|i-j|/\xi}$.
The optimal hopping range~\cite{Mott1969} of a fast electron is then controlled by the competition between this decay and the higher probability of reaching a hole with high phonon number, which increases $D_{nm}$.
Therefore, we maximise 
\begin{equation}
    \Gamma^*_n = t e^{-R/\xi} |D_{nm(R)}|^2,
\end{equation}
where $m(R)$ is chosen such that the probability of having a hole with phonon number greater than $m$ within range $R$ is a fixed finite value (for simplicity, we choose $1-1/e$):
\begin{equation}
    (1-e^{-\beta\Omega m(R)})^{2R} = e^{-1} \implies m(R) \approx \frac{\ln (2R)}{\beta\Omega}.
\end{equation}
Performing the saddle-point optimisation (see~\cite{supplement} for details) we find an optimal hopping range $R^*\sim \xi/\beta\Omega$ with a self-averaging distribution of $R$ around it, except for extremely fast electrons. Namely, if $A^2-n < A$, the hopping rate decreases for all positive $R$, i.e., nearest-neighbour hops dominate. 
In this limit, we find
\begin{equation}
    \Gamma^*_n \approx t e^{-A^2} \frac{A^{2n}}{n!}.
    \label{eq: Gamma very fast electron}
\end{equation}

Finally, we would like to quantify which electrons are ``fast'' and ``slow,'' that is, which ones are efficient at scrambling the phonon occupation numbers of all sites. 
The rate at which a randomly chosen empty site meets a electron of phonon number $n$ is the product of the hopping rate $\Gamma^*_n$ and their density, proportional to the Boltzmann factor $e^{-\beta\Omega n}$. The total scrambling rate is thus
\begin{equation}
    \Gamma_0 = \rho(1-e^{-\beta\Omega}) \int \ud n \Gamma^*_n e^{-\beta\Omega n}.
    \label{eq: saddle point thermal}
\end{equation}
Again estimating the integral by  saddle-point~\cite{supplement},
we find that, over a wide temperature range $U_\mathrm{e-ph} \gg T \gtrsim A\Omega$, \eqref{eq: saddle point thermal} is dominated by electrons of extremely high phonon numbers $n^*\approx A^2 e^{-\beta\Omega}$;
\eqref{eq: Gamma very fast electron} then yields
\begin{equation}
   \!\! \Gamma_{0,h} \approx \rho(1-e^{-\beta\Omega}) t e^{n^*-A^2} \approx \rho(1-e^{-\beta\Omega}) t e^{-U_\mathrm{e-ph} / T}.
\end{equation}
Repeating the same calculation for fast holes, we find
\begin{equation}
    \Gamma_{0,e} \approx (1-\rho)(1-e^{-\beta\Omega}) t e^{-U_\mathrm{e-ph} / T}
\end{equation}
to be the mean hopping rate of slow electrons.

Low-frequency conductivity in our model is thus dominated by faciliated diffusive processes in which ``slow'' holes (i.e., ones with $n\ll n^*$) swap places with ``fast'' electrons (i.e., ones with $n\approx n^*$), or vice versa.
One can account for such processes keeping track of any of the species involved; we choose to focus on the slow particles, for they make up the vast majority of available carriers and their hopping rate is uniformly $\Gamma_0$. 
Their movement is entirely due to ``fast'' particles, so it can be treated as an incoherent, classical random walk between nearest neighbours at rate $\Gamma_0$. Such a process yields the dc conductivity~\cite{Mott1969,Austin1969}
\begin{equation}
    \sigma_\mathrm{dc} = \beta e^2 L_0[\rho \Gamma_{0,f} + (1-\rho)\Gamma_{0,h}] \approx \sigma_0 e^{-U_\mathrm{e-ph}/T}.
\end{equation}
We also expect that this is a good approximation of the ac conductivity for $\omega\ll\Gamma_0$, since the time dependence of the external electric field is slow compared to the hopping of carriers, and the fast scrambling of sites also disrupts the coherent two-site processes discussed above.
At the highest frequencies, however, the latter are expected to take over and dominate incoherent transport, since $\sigma_\mathrm{max} = \sigma_0 e^{-U_\mathrm{e-ph}/2T} \gg \sigma_\mathrm{dc}$. At $\omega\approx \Gamma_0$, multi-site coherent and two-site incoherent~\cite{Austin1969} processes may also contribute significantly to $\sigma$; treating these in full detail is beyond the scope of this paper.

\textit{Behaviour at finite $t$.}---%
So far, we only considered first-order effects of electron hopping.
In second-order perturbation theory, virtual hopping processes introduce self-energy corrections to the energies of the states $|\dot m n\rangle$ and $|n\dot m\rangle$ in the coherent-hopping scenario above. Such corrections are dominated by nearby fast (in the sense described above) electrons and holes, which enhance FC factors to order unity.
As a result, energies of the supposedly degenerate states $|\dot m n\rangle$ and $|n\dot m\rangle$ may randomly be detuned by $\sim t^2/\hbar|\Omega'-\Omega|$, where $\Omega'$ is the phonon frequency on occupied sites; $\Omega'-\Omega\sim cU_\mathrm{e-ph}/\sqrt{k^3m}$ due to anharmonicity.
This is enough to make the transition between them off-resonant for low phonon occupation such that $D_{nm} \lesssim t/|\Omega'-\Omega|$. 
For small enough $t$, this only enhances the disruption of coherent processes at low frequencies; if $t\sim |\Omega'-\Omega|$, however, no processes of the kind discussed earlier survive. 
Indeed, the emergent phonon-number conservation underpinning our perturbative treatment breaks down, as the kinetic energy of electrons can make up for the energy differences between phonon states. If $|\Omega'-\Omega|\ll\Omega$, we enter a regime where phonon number is conserved overall, but can be freely rearranged between sites: understanding this regime (or, equivalently, the case of harmonic phonons) is an intriguing if challenging direction for the future.

\textit{Conclusion.---}%
We explored high-temperature transport in the anharmonic Holstein model, focusing on the strong-coupling limit.
We argued that phonon anharmonicity introduces an emergent conservation law, which attaches a phonon occupation number to all electrons and holes that remains unchanged in hopping processes: a process we term `polaron speciation'.
Our work exemplifies  unusual  transport phenomena that  can be uncovered  even in relatively well-studied models of condensed matter by exploring poorly-explored regimes of parameters, in this case high temperature and strong coupling. Understanding the dynamics in these extreme regimes benefits from ideas and techniques sharpened by recent forays into even more exotic settings, such as MBL. Encouragingly, our results  suggest that systems at the threshold of localization  may offer new solid-state experimental platforms for investigating unconventional quantum dynamics. It would be particularly interesting to explore these regimes in the setting of narrow bands engineered in moir\'e superlattices, where the role of electron-phonon interactions remains debated~\cite{Lian2019}.

\vspace{0.2cm}

\textit{Note added.---} As this paper was being finalised, we became aware of complementary work by Esterlis, Murthy, and Kivelson that addresses the perfectly-static phonon regime accessed by taking the limit of infinite phonon mass. Our results agree where they overlap.

\vspace{0.3cm}

\noindent\textit{Acknowledgements.---}%
S.\ G. thanks C. Murthy for helpful discussions. We acknowledge support from the Oxford--ShanghaiTech collaboration project (A.\ Sz.) and from the UK Engineering and Physical Sciences Research Council via EPSRC grant EP/S020527/1. Statement of compliance with EPSRC policy framework
on research data: This publication is theoretical work
that does not require supporting research data.

\bibliography{paper}

\end{document}


\title{Supplementary material to ``Finite-temperature transport and polaron speciation in the anharmonic Holstein model''}
\author{Attila Szabó}
\address{Rudolf Peierls Centre for Theoretical Physics, University of Oxford, Oxford OX1 3PU, United Kingdom}
\address{ISIS Facility, Rutherford Appleton Laboratory, Harwell Campus, Didcot OX11 0QX, UK}
\author{S. A. Parameswaran}
\address{Rudolf Peierls Centre for Theoretical Physics, University of Oxford, Oxford OX1 3PU, United Kingdom}
\author{Sarang Gopalakrishnan}
\affiliation{Department of Physics, The Pennsylvania State University, University Park, PA 16802, USA}

\date{\today}

\maketitle

\section{Asymptotes of the Franck--Condon factors}

We would like to calculate the generalised Franck--Condon factor $D_{nk} = \langle n| k'\rangle $ in the strong-coupling limit, where $|n\rangle$ and $|k'\rangle$ are phonon eigenstates for empty and occupied sites, respectively.
We work in the approximation that phonons on both empty and occupied sites are described by harmonic-oscillator Hamiltonians, but their respective frequencies, $\Omega$ and $\Omega'$, may differ:
\begin{align}
    H_\mathrm{empty} &= \frac{p^2}{2m} + \frac{m\Omega^2}2 x^2; &
    H_\mathrm{occupied} &= \frac{p^2}{2m} = \frac{m\Omega'^2}2 (x-x_0)^2.
\end{align}
In particular, we look for an asymptotic form for large $n$ and $k$, which we obtain by saddle-point approximating the power series
\begin{equation}
    \sum_{n,k=0}^{\infty} D_{nk} \frac{\alpha^n}{\sqrt{n!}}\frac{\beta^k}{\sqrt{k!}} = e^{(|\alpha|^2+|\beta|^2)/2} \langle \bar\alpha | \beta'\rangle,
\end{equation}
where $|\alpha\rangle,|\beta'\rangle$ are coherent states for empty and occupied sites. These coherent states can be written in position basis as
\begin{align}
    e^{|\alpha|^2/2}\langle \bar\alpha|x\rangle  = e^{|\alpha|^2/2}\langle x | \alpha\rangle &=  \left(\frac{m\Omega}\pi\right)^{1/4}  \exp\left[-\left(\sqrt{\frac{m\Omega}{2}}x-\alpha\right)^2 + \frac{\alpha^2}2\right]\\
    e^{|\beta|^2/2} \langle x |\beta'\rangle &= \left(\frac{m\Omega'}\pi\right)^{1/4}  \exp\left[-\left(\sqrt{\frac{m\Omega'}{2}}(x-x_0)-\beta\right)^2 + \frac{\beta^2}2\right],
\end{align}
whence
\begin{align}
    e^{(|\alpha|^2+|\beta|^2)/2} \langle \bar\alpha | \beta'\rangle &= \sqrt{\frac{m\sqrt{\Omega\Omega'}}\pi} \int_{-\infty}^\infty \ud x \exp\left[\frac{\alpha^2+\beta^2}2 - \left(\sqrt{\frac{m\Omega}{2}}x-\alpha\right)^2-\left(\sqrt{\frac{m\Omega'}{2}}(x-x_0)-\beta\right)^2 \right]  \nonumber\\
    &= \underbrace{\sqrt{\frac{2\sqrt{\Omega\Omega'}}{\Omega+\Omega'}}}_X \exp\underbrace{\left(\frac{-mx_0^2\Omega\Omega' +\sqrt{8m\Omega\Omega'}x_0(\alpha\sqrt{\Omega'}-\beta\sqrt\Omega) +(\alpha^2-\beta^2)(\Omega-\Omega') + 4\alpha\beta\sqrt{\Omega\Omega'} }{2(\Omega+\Omega')} \right)}_{I}.
    \label{eq: power series result}
\end{align}
Now, $D_{nk}$ can be obtained from the integral
\begin{equation}
    D_{nk} = \frac{X\sqrt{n!k!} }{(2\pi i)^2} \oint \ud\alpha\ud\beta \frac{e^{I} }{\alpha^{n+1} \beta^{k+1}},
    \label{eq: contour integral}
\end{equation}
where the integral runs over a torus in $(\alpha,\beta)$ space that winds once around the origin in both variables. The asymptotic behaviour of $D_{nk}$ is controlled by the stationary points of the integrand (or, equivalently, its log):
\begin{subequations}
\label{eq: saddle point general}
\begin{align}
    \frac{\partial}{\partial\alpha}\left(I-n\log\alpha - k\log\beta\right) = \frac{\partial I}{\partial\alpha} &-\frac{n}{\alpha} = 0 \nonumber\\
    \frac{(\Omega-\Omega')\alpha^2 + 2\sqrt{\Omega\Omega'}\alpha\beta + \sqrt{2m\Omega}x_0\Omega'\alpha}{\Omega+\Omega'} &= n;\\
    \frac{\partial}{\partial\beta}\left(I-n\log\alpha - k\log\beta\right) = \frac{\partial I}{\partial\beta} &-\frac{n}{\beta} = 0 \nonumber\\
    \frac{(\Omega'-\Omega)\beta^2 +2\sqrt{\Omega\Omega'}\alpha\beta -\sqrt{2m\Omega'}x_0\Omega\beta}{\Omega+\Omega'} &= k.
\end{align}
\end{subequations}
In general, solving~\eqref{eq: saddle point general} leads to a quartic equation, so it can at best be done numerically. However, for the case of standard Franck--Condon factors, $\Omega'=\Omega$, the $\alpha^2-\beta^2$ term disappears from~\eqref{eq: power series result}, which makes the saddle-point equations quadratic. It also makes sense to introduce the dimensionless shift $A = x_0\sqrt{m\Omega/2}$, in terms of which we get the equations
\begin{align}
    n &= \alpha\beta + A\alpha; & k &= \alpha\beta - A\beta.
\end{align}
This is solved by
\begin{align}
    \alpha &=\frac{A^2+n-k\pm\sqrt{A^4 - 2A^2(k+n) +(k-n)^2}}{2A}; &
    \beta &= -\frac{A^2+k-n\pm\sqrt{A^4 - 2A^2(k+n) +(k-n)^2}}{2A};
    \label{eq: saddle point solution}
\end{align}
which we plug back into the simplified form of~(\ref{eq: power series result},\ref{eq: contour integral}),
\begin{align}
    I &=  -A^2 + A(\alpha-\beta) + \alpha\beta = \frac{k+n \pm \sqrt{A^4 - 2A^2(k+n) +(k-n)^2}}2\nonumber\\
    \log|D_{nk}| &\approx \frac12(n\log n-n+k\log k-k) +I - n\log|\alpha| - k\log|\beta|.
    \label{eq: simplified FC factor}
\end{align}
To choose between the positive and negative roots in~\eqref{eq: saddle point solution}, we consider $n=k=0$, where the overlap is easily shown to be $e^{-A^2/2}$: \eqref{eq: simplified FC factor} yields $e^{\pm A^2/2}$ for the two different solutions, so we should pick the one with the minus sign. Hence, we get $\log | D_{nk} | = -A^2 F(n/A^2,k/A^2)$, where 
\begin{equation}
    F(x,y) = \frac{\sqrt D}2 + 
    x \log \frac{1+(x-y)-\sqrt D}{2} -\frac{x}2 \log x + 
    y \log \frac{1+(y-x)-\sqrt D}{2} -\frac{y}2 \log y
    \label{eq: FC factor asymptote}
\end{equation}
and $D = 1-2(x+y)+(x-y)^2$.

\section{Scaling of the two-site conductivity}

As discussed in the main text, the contribution of two-site coherent processes to ac conductivity is given by
\begin{align}
    \sigma_{ij}(\omega) &= 2\pi\beta L \frac{e^2}{\hbar} \rho(1-\rho)(1-e^{-\beta\hbar\Omega})^2  \sum_{n,m=0}^\infty e^{-\beta\hbar\Omega(n+m)}t_{ij}^2 |D_{mn}|^4  \delta(\hbar\omega - 2t_{ij} |D_{mn}|^2)\nonumber\\
    &= \frac{\pi\beta L e^2\hbar}2 \rho(1-\rho)(1-e^{-\beta\hbar\Omega})^2 \omega^2\sum_{n,m=0}^\infty e^{-\beta\hbar\Omega(n+m)} \delta(\hbar\omega - 2t_{ij} |D_{mn}|^2);
    \label{eq: two state Kubo formula}
\end{align}
we now evaluate this sum to leading exponential order, indicating how one could obtain the pre-exponential factors if needed.

First, we observe that our asymptote of the Franck--Condon factor $D_{nm}$, Eq.~\eqref{eq: FC factor asymptote}, is not written directly in terms of $D$ but its logarithm. Therefore, we perform the following change of variables:
\begin{align}
    \delta(\hbar\omega - 2t_{ij}|D_{nm}|^2) &= \frac1{\hbar\omega}\, \delta\left(\log \frac{\hbar\omega}{2t_{ij}|D_{nm}|^2}\right) = \frac1{\hbar\omega}\  \delta\left(\log\frac{\hbar\omega}{2t_{ij}} - 2\log |D_{nm}|\right), \nonumber
\end{align}
whence~\eqref{eq: two state Kubo formula} can be rewritten as
\begin{equation}
    \sigma(\omega) = \frac{\pi\beta L e^2}2 \rho(1-\rho)(1-e^{-\beta\hbar\Omega})^2 \omega \sum_{n,m=0}^\infty e^{-\beta\hbar\Omega(n+m)}\ \delta\left(\log\frac{\hbar\omega}{2t_{ij}} - 2\log |D_{nm}|\right).
    \label{eq: two site conductivity no approx}
\end{equation}
Now, assuming $\Omega\ll T$, the exponential factors inside the sum change slowly enough in $n,m$ that the sum can be replaced by an integral, and we can substitute the asymptote~\eqref{eq: FC factor asymptote} for $\log |D_{nm}|$. 
The $\delta$-function reduces the resulting surface integral to a line integral of $e^{-\beta\hbar\Omega(n+m)}$, corrected by the derivative of the argument of the $\delta$-function. 
Since the latter is always $O(A^2)$, it does not give rise to exponential corrections, so we can neglect it to leading order in the saddle-point approximation.
This leaves us with finding the lowest $n+m$ for a given $|D_{nm}|$ or, equivalently, the highest $|D_{nm}|$ for a given $n+m$.
By calculating the first two derivatives of $\log|D_{n,N-n}|$ with respect to $n$ from the asymptotic form~\eqref{eq: FC factor asymptote}, we find that this occurs at $n=m$, whence
\begin{equation}
    \sigma(\omega) \sim \sigma_0 e^{-2\beta\hbar\Omega \bar n} t_{ij} |D_{\bar n\bar n}|^2,
\end{equation}
where $\bar n$ is such that $\hbar\omega = 2t_{ij}|D_{\bar n\bar n}|^2$;
we lump all constant prefactors into a $T$-dependent $\sigma_0$,
and neglect pre-exponential factors from the $\delta$-function and the saddle-point integral.

Given the complicated expression for $D_{nn}$, we cannot write $\sigma$ explicitly in terms of $\omega$: instead, we express it implicitly in terms of $x = n/A^2$ as
\begin{subequations}
\label{eq: two site conductivity}
\begin{align}
    \log(\hbar\omega/2t_{ij}) &= 2\log|D_{nn}| = A^2\left(\sqrt{1-4x} + 2x\log \frac{1-\sqrt{1-4x}}2 -x\log x\right);
    \label{eq: two site conductivity omega}\\
    \log(\sigma/\sigma_0) &= -2\beta\hbar\Omega n + 2\log|D_{nn}| = A^2\left( \sqrt{1-4x} + 2x\log \frac{1-\sqrt{1-4x}}2 -x\log x - 2\beta\hbar\Omega x\right),
    \label{eq: two site conductivity sigma}
\end{align}
\end{subequations}
which is plotted in the main text.

\section{Saddle-point optimisation of the incoherent hopping rate}
\label{appendix: hopping rate saddle point}

\subsection{Fixed phonon number on the electron site}
\label{appendix: fixed phonon number saddle point}

We first maximise $\Gamma = t e^{-R/\xi} |D_{nm(R)}|^2$:
\begin{align}
    0=\frac{\ud}{\ud R} \left(-\frac{R}{\xi} + 2 \log |D_{nm(R)}| \right) = -\frac1\xi + 2\frac{\ud m}{\ud R}\frac{\ud\log |D_{nm}|}{\ud m} = -\frac1\xi - \frac{2}{\beta\Omega R} F^{(0,1)}(n/A^2,m(R)/A^2),
\end{align}
where $F$ is the asymptote given in~\eqref{eq: FC factor asymptote}. Assuming the optimal $R$ is on the order of $\xi$, we expect that the optimal $m(R)$ is much smaller than $A^2$, so we can expand $F$ in small second arguments:
\begin{align}
    F^{(0,1)}(x,y)|_{y\to 0} &\approx \frac{\log y}2 - \log(1-x)\\
    \beta\Omega R &\approx \xi \left[ -\log\frac{m}{A^2} +2\log \left(1-\frac{n}{A^2}\right) \right].
\end{align}
Again, we neglect $\log m$ in comparison with $\log A^2$,%
\footnote{The small-$y$ expansion of $F$ itself shows that the Franck--Condon factor scales approximately as $|A|^{m}/\sqrt{m!}$ (this could also be seen from the appropriate Laguerre polynomial). Neglecting $\log m$ is equivalent to dropping the factorial. \label{footnote: FC factor approx}} 
which leads to
\begin{align}
    R^* &= \frac{\xi \log[(A^2-n)^2/A^2]}{\beta\Omega}; &
    m^* &= \frac1{\beta\Omega} \log\left(\frac{2\xi \log[(A^2-n)^2/A^2]}{\beta\Omega}\right).
    \label{eq: optimal R, m}
\end{align}
In the latter, the only term that doesn't enter logarithmically is $\beta\Omega$; therefore, $m\ll A^2$ is a good approximation as long as $1/(\beta\Omega) \ll A^2 = U_\mathrm{e-ph}/\Omega$, i.e., $T\ll U_\mathrm{e-ph}$, the electron--phonon interaction scale.

The distribution of $R$ around $R^*$ is self-averaging. The log of the hopping rate as a function of $R$ for $m\ll A^2$ is
\begin{align}
    \log(\Gamma/t) = -\frac{R}\xi + 2 \log |D_{nm(R)}| &\approx -\frac{R}\xi -A^2 +(n+m)\log A^2 - (\log n! + \log m!) + 2m \log \left(1-\frac{n}{A^2}\right) \nonumber\\
    &\approx -\frac{R}{\xi} + m\log[(A^2-n)^2/A^2] + \left\{n\log A^2 - \log n! - A^2\right\}
    \label{eq: hopping rate}\\
    \Gamma &\approx \mathrm{const.}\times e^{-R/\xi} (2R)^{\log[(A^2-n)^2/A^2] / \beta\Omega}.
\end{align}
We see the same integrand in the standard derivation of Stirling's formula for $\Gamma\big\{\log[(A^2-n)^2/A^2]/\beta\Omega\big\}$: as long as the argument is large (this follows in the strong-coupling limit as long as $T\gg\Omega$, which we need to treat phonon numbers as a continuum anyway), the saddle-point approximation to this consists of a sharp peak, well approximated by a Gaussian.

There is one exception to this behaviour. If $A^2-n < A$, the log in the result~\eqref{eq: optimal R, m} for $R^*$ becomes negative, which implies that the hopping rate decreases for all positive range $R$, i.e., the optimal hopping range $R^*$ is to nearest neighbours. This edge case will turn out to be important when optimising over~\eqref{eq: hopping rate} later.

\subsection{Thermal ensemble of polarons}
\label{appendix: thermal saddle point}

We now want to calculate the rate at which the whole system scrambles, i.e., at which even the slowest electrons/holes participate in a hopping process due to their faster neighbours. 
This we can do by integrating over the rate~\eqref{eq: hopping rate} found above, multiplied by the Boltzmann factor $e^{-\beta\Omega n}$: this is the rate at which a randomly picked site will be perturbed by a fermion with occupation number $n$.

We perform this integral in the saddle-point approximation too. Plugging~\eqref{eq: optimal R, m} into~\eqref{eq: hopping rate} gives
\begin{align}
    \log(\Gamma^* e^{-\beta\Omega n}/t) \approx -A^2 + \frac{\log[(A^2-n)^2/A^2]}{\beta\Omega} \left(\log\frac{2\xi\log[(A^2-n)^2/A^2]}{\beta\Omega}-1\right) +n \log A^2 - \log n! - \beta\Omega n.
    \label{eq: hopping rate with boltzmann}
\end{align}
The $n$ that contributes the most to the scattering rate is found by setting the derivative of this to 0:
\begin{align}
    0 &= -\frac{2}{\beta\Omega(A^2-n)} \log\frac{2\xi\log[(A^2-n)^2/A^2]}{\beta\Omega}  +\log A^2 - \log n - \beta\Omega \nonumber\\
    \frac{2\xi m^*}{A^2-n} + \beta\Omega &=  \log\frac{A^2}n.
    \label{eq: saddle point for n}
\end{align}
Let us first focus on the regime $T\gtrsim A\Omega$ and forget about the first term for a moment. 
Then, we get $n= A^2 e^{-\beta\Omega}\approx A^2 - A^2\beta\Omega$: the second term is smaller than $A$ by construction, so we are in the limit $A^2-n < A$ flagged up at the end of \S\ref{appendix: fixed phonon number saddle point}. This means that the saddle-point calculation in the same section collapses, and we find that the optimal hopping partner is the nearest neighbour of the fast fermion, regardless of its occupation number. This eliminates the complicated second term from the right hand side of~\eqref{eq: hopping rate with boltzmann}, as well as the first term of~\eqref{eq: saddle point for n}. That is, $n^* = A^2 e^{-\beta\Omega}$ solves the saddle-point equation exactly.

Furthermore, even if $T<A\Omega$, the first term of~\eqref{eq: saddle point for n} would be parametrically small (in the same approximation, we would have $A^2-n > A \gg m^*$), so to a good first approximation, we would always (except for very small temperatures maybe) have
\begin{equation}
    n^* = A^2 e^{-\beta\Omega}.
\end{equation}
Plugging this into~\eqref{eq: hopping rate with boltzmann} without the second term yields
\begin{equation}
    \Gamma^* = t e^{-(A^2-n)} \approx t e^{-A^2\beta\Omega}  = t e^{-U_\mathrm{e-ph}/T}.
\end{equation}